%% file: main.tex
\documentclass[twoside,leqno,twocolumn]{article}

\usepackage{siamproceedings}
\usepackage{fancyhdr}

\title{\replaced[id=R5]{Compiler-supported}{Combining} reduced precision and AoS-SoA transformations \replaced[id=R5]{for}{to accelerate SPH calculations on} \replaced[id=Ours]{heterogeneous}{heterogenous} hardware}
\date{}
\author{
 Pawel K.~Radtke\thanks{
  Department of Computer Science, Institute for Data Science, Durham University,
  United Kingdom
  \href{mailto:pawel.k.radtke@durham.ac.uk}{pawel.k.radtke@durham.ac.uk}
  }
 \and
 Tobias Weinzierl\thanks{
  Department of Computer Science, Institute for Data Science, Durham University,
  United Kingdom
  \href{mailto:tobias.weinzierl@durham.ac.uk}{tobias.weinzierl@durham.ac.uk}
}
}

\usepackage{algorithm}
\usepackage{algorithmicx}
\usepackage{algpseudocode}
\usepackage{amsfonts}
\usepackage{amsmath}
\usepackage{cite}
\usepackage{float}
\usepackage{graphicx}
\usepackage{subcaption}
\usepackage{url}

\usepackage[final,commandnameprefix=ifneeded]{changes}
 
\definechangesauthor[name=Review1, color=blue]{R1}
\definechangesauthor[name=Review2, color=red]{R2}
\definechangesauthor[name=Review3, color=green]{R3}
\definechangesauthor[name=Review4, color=purple]{R4}
\definechangesauthor[name=Review5, color=orange]{R5}
\definechangesauthor[name=Ours, color=teal]{Ours}
\definechangesauthor[name=HT, color=yellow]{HT}

\usepackage[a4paper, total={7in, 10in}]{geometry}

\begin{document}

\maketitle

\begin{abstract}
\input{00_abstract.tex}
\end{abstract}


\section{Introduction}
\input{01_intro}

\section{Smoothed Particle Hydrodynamics}
\input{02_sph}

\section{Source code annotations}
\input{03_cpp_extensions}

\section{Formal code transformations and algorithm description}
\input{03_formalisms}

\section{Compiler implementation}
\input{04_compiler_implementation}

\section{Results}
\input{05_results}

\section{\added[id=R1]{Numerical impact of floating-point truncation}}

\input{08_correctness}

\section{Conclusion}
\input{09_conclusion}

\section*{Acknowledgements}
\input{99_acknowledgements}

\bibliography{paper}
\bibliographystyle{IEEEtran} 

\appendix
\section{Compiler prototype}
\input{appendix_clang}

\end{document}

%% file: 00_abstract.tex
This study evaluates AoS-to-SoA transformations \replaced[id=R4]{over}{and} reduced-precision data layouts \deleted[id=R4]{across multiple compute kernels} for a particle simulation code on several GPU platforms:
We hypothesize that SoA  \replaced[id=R4]{fits particularly well to SIMT}{facilitates aggressive vectorisation}, while AoS is the preferred storage format for many Lagrangian codes\replaced[id=R4]{. Reduced-precision}{ and reduced-precision data storage} (below IEEE accuracy) \replaced[id=R4]{is an established tool}{for some quantities helps us} to address bandwidth constraints\replaced[id=R4]{, although}{.
Therefore, it is sometimes beneficial to convert data pior to their first usage, i.e., on-the-fly into SoA over native data formats.
We introduce C++ annotations to facilitate such a conversion.
However,} \replaced[id=R4]{it remains unclear whether AoS and precision conversions}{not clear if such a conversion} should \replaced[id=R4]{execute}{be done} on a CPU or be deployed to a GPU if the compute kernel itself \replaced[id=R4]{must}{has to} run on an accelerator.
\added[id=R4]{
  On modern superchips where CPUs and GPUs share (logically) one data space,
  it is also unclear whether it is advantageous to stream data to the accelerator prior to the calculation,
  or whether we should let the hardware transfer data transparently on demand, i.e.~work in-place logically.
}
We \added[id=R4]{therefore} introduce compiler annotations to facilitate such \replaced[id=R4]{conversions and to give the programmer the option to orchestrate the conversions in combination with GPU offloading.}{a conversion.}
\deleted[id=R4]{Performance gains depend on kernels' arithmetic intensity, memory access patterns, and hardware characteristics. Memory-bound kernels achieve the highest speedups from data format rearrangements and reduced precision storage, while compute-intensive kernels show multifacted behaviour - an SPH density kernel exhibits platform-dependent results, i.e., performs well on Nvidia hardware but inconsistently on AMD. 
In the best case,}
\added[id=R4]{For some of our compute kernels of interest,} Nvidia's G200 platforms yield a speedup of around 2.6\replaced[id=R4]{
 while AMD's MI300A exhibits more robust performance yet profits less.
 We assume that our compiler-based techniques are applicable to a wide variety of Lagrangian codes and beyond. 
}{using GPU-based SoA conversion and reduced-precision. 
AMD systems showed constrained benefits, particularly for larger working sets, with maximum gains of 1.4 for 16-bit floating-point storage.}

%% file: 01_intro.tex
The move into the exascale era has shifted the dominant \added[id=Ours]{runtime} constraint from floating-point throughput to data movement~\cite{Dongarra:2011:ExascaleSoftwareRoadmap,McCalpin:1995:Stream}. Modern accelerators can sustain very high \replaced[id=Ours]{throughput}{single-precision rates}, yet the bandwidth and latency of moving data to and from compute units often govern end-to-end performance. Typical workloads illustrate this imbalance: memory-bound kernels with low arithmetic intensity saturate device memory channels long before the ALUs are fully utilised; fine-grained kernels pay disproportionate costs for host–device transfers and kernel-launch overheads; and irregular access patterns defeat caching and prefetchers~\cite{Che:2011:Dymaxion,Majeti:2014:CompilerDrivenDataLayoutTransformations}. Energy consumption follows the same pattern: moving data across memory hierarchies and interconnects consumes more energy per element than performing simple arithmetic on it~\cite{Tsai:2019:CompressObjects}. Reduced precision has therefore become a practical lever—decreasing transfer volume and increasing effective bandwidth—but its safe use is application-dependent and typically requires explicit control over where precision can be traded for speed without compromising numerical \replaced[id=Ours]{accuracy}{objectives}~\cite{Nasar:2024:GPU,Schaller:2024:Swift}.

Programming models and frameworks address pieces of this problem, but rarely the whole. 
CUDA, HIP, SYCL, and OpenMP offload expose accelerators, asynchronous execution, streams/queues, and explicit memory management~\cite{OpenMP:TechnicalReport8}. 
Recent features such as \deleted[id=Ours]{graphs, cooperative groups,} unified/managed memory\replaced[id=Ours]{ as provided in hardware on modern superchips such as Nvidia's Grace Hopper or AMD's MI300a, while they previously had to be provided in software by the runtime,}{,} and memory pools reduce overheads or programmer burden. Libraries and portability layers such as Kokkos and RAJA provide abstractions for execution spaces and views, and encourage layout-aware programming~\cite{KOKKOS, Trott:2022:Kokkos, RAJA}. In practice, however, developers still decide when and how to transform data layouts~\cite{Radtke:2024:AoS2SoA,Homann:2018:SoAx}, where to place \replaced[id=Ours]{conversions}{packing and unpacking work} (CPU or GPU)~\cite{Che:2011:Dymaxion,Sung:2012:DataLayoutTransformations}, and how to couple layout with precision policy. Mixed-precision support is strong in dense linear-algebra libraries and tensor-core paths, but general scientific codes often rely on hand-crafted conversions around application kernels~\cite{Gallard:2020:Roles,Gallard:2020:ExaHyPEVectorisation}. 
As a result, \replaced[id=Ours]{precision remains another}{memory layout and precision remain} first-order \replaced[id=Ours]{performance knob that is only}{performance knobs that are only} weakly captured by current language and framework mechanisms~\cite{Majeti:2014:CompilerDrivenDataLayoutTransformations}.

Smoothed Particle Hydrodynamics (SPH) exemplifies these issues in a scientific-computing context. SPH represents fluids as a set of particles carrying physical quantities such as position, velocity, and density, with interactions computed through kernel-weighted sums over particle neighbourhoods~\cite{Monaghan:1985:Kernel,Price:2012:SPH,Lind:2020,Schaller:2024:Swift}. This leads to two distinct classes of compute kernels: linear kernels, where each particle updates its state independently, and quadratic kernels, where contributions arise from particle–particle interactions. The latter dominate the runtime due to their $\mathcal{O}(N \cdot N_{ngb})$ complexity, with $N_{ngb}$ denoting the approximately constant number of neighbours per particle. Despite being largely local, these operations are difficult to accelerate effectively~\cite{Nasar:2024:GPU,Wille:2023:GPU}. 
Most SPH implementations store particle data in an array-of-structs (AoS) layout that is natural for software engineering but poorly matched to accelerator memory systems, resulting in non-coalesced accesses and inflated transfer sizes~\cite{Homann:2018:SoAx,Radtke:2024:AoS2SoA,Slattery:2022:Cabana}. 
Furthermore, SPH codes typically \added[id=Ours]{can} employ single \added[id=Ours]{or smaller} precision\replaced[id=Ours]{---at least for some particle properties}{to balance accuracy and performance}, avoiding the double precision throughput penalty\deleted[id=Ours]{ but still leaving bandwidth and transfer efficiency as limiting factors~\cite{McCalpin:1995:Stream}}.
Together, these characteristics make SPH a natural \replaced[id=Ours]{candidate}{case study} for exploring data-layout and reduced-precision strategies on heterogeneous architectures~\cite{Dongarra:2011:ExascaleSoftwareRoadmap}.

This work addresses \replaced[id=Ours]{the arising implementation}{these} challenges through a combination of language-level extensions\replaced[id=Ours]{ and}{,} compiler support\deleted[id=Ours]{, and systematic evaluation}. 
First, we introduce C++ annotations that allow developers to request temporary, on-the-fly memory-layout transformations \deleted[id=Ours]{and reduced-precision storage} directly in the source code. These annotations enable automatic array-of-structs to struct-of-arrays (AoS$\mapsto $SoA) conversions that are specialised to the needs of individual kernels. 
Second, we study two complementary strategies for constructing \added[id=Ours]{these} SoA views: 
performing the conversion on the CPU before transferring data to the GPU, \replaced[id=Ours]{or}{and} transferring AoS buffers directly to the GPU where the conversion is executed\deleted[id=Ours]{ in parallel}. 
\added[id=Ours]{Finally, we facilitate reduced-precision storage of attributes.}
All three mechanisms are realised transparently through compiler transformations that integrate with existing OpenMP-offload paths, requiring no manual \replaced[id=Ours]{intervention}{packing code} from the application developer. 
\replaced[id=Ours]{We}{Finally, we} evaluate these techniques on four representative accelerator systems—two PCIe-attached devices and two superchip platforms—to quantify their impact on both computation and data movement.

\added[id=R1]{Our prior work already introduced local AoS$\mapsto $SoA conversions implemented via C++ attributes~\cite{Radtke:2024:AoS2SoA}, including support for heterogeneous hardware~\cite{RadtkeWeinzierl:CCPEAoSSoA}. 
In the present paper, we investigate host-driven and accelerator-driven AoS$\mapsto $SoA conversion strategies, and combine them with reduced-precision storage as championed in~\cite{Radtke:2027:CPP}. 
We also present performance measurements of kernel-level microbenchmarks, as well as whole-timestep performance based for two different memory-management approaches: 
in-place conversion, where compressed variables are mapped from and to native hardware formats immediately prior or after each access but the data resides on the host even though we offload to an accelerator,
or streaming-based, where we copy rearranged data en bloc over to the GPU prior to invoking a compute kernel.
}
\added[id=R2]{Not provided is an investigating the hardware behaviour that underpins and explains the performance characteristics of our proposed contributions. 
Given that we observe differences across hardware vendors, a dedicated study of vendor-specific behaviour and possible optimisations has to be part of future work,
i.e.~the techniques themselves are of value---also outside of an SPH/Lagrangian regime---but impact statements have to depend strongly on a better understanding of the influence of hardware characteristics. 
}

The remainder of this paper is organised as follows. Section 2 introduces Smoothed Particle Hydrodynamics (SPH) as our case study, outlining its governing equations and computational structure. Section 3 presents the language extensions that allow developers to annotate data-layout and precision requirements, while Section 4 describes the underlying formalisms. Section 5 details the compiler implementation that realises these annotations as transparent transformations. Section 6 presents the experimental results, including the hardware platforms and benchmarking setup, and Section 7 concludes with a summary of contributions and directions for future research.

%% file: 02_sph.tex
\added[id=R5]{
Our work extends the C++ language in an application domain-agnostic way. 
Although we use Smoothed Particle Hydrodynamics as a demonstrator---both because it traditionally favours AoS and because its kernels clearly expose the limitations of current language and compiler support---SPH serves only as motivation. 
All mechanisms as introduced do not rely on any SPH-specific semantics, data structures, or computational patterns. They apply equally to any domain where structured aggregates are traversed repeatedly, transformed into more cache- and SIMT-friendly layouts, or offloaded to heterogeneous hardware.
}

\subsection{Governing equations}

SPH is a mesh-free method that represents a fluid as a set of particles carrying mass, momentum, energy, and other physical quantities~\cite{Monaghan:1985:Kernel,Price:2012:SPH,Lind:2020}. Each particle interacts with its neighbours through a smoothing kernel function. This kernel interpolates a continuous field quantity $A(r)$ from discrete particle data as follows~\cite{Springel-g2:2005,Schaller:2024:Swift}:

\begin{equation}
A(r_i) \approx \sum_{j} \frac{m_j}{\rho_j} A_j W(r_i - r_j, h_{ij})
\end{equation}

where $r_j$, $m_j$, and $\rho_j$ denote the position, mass, and density of a neighbouring particle $j$, and $h_{ij}$ is a linear combination of smoothing lengths $h_i$ and $h_j$. The smoothing kernel $W(r, h)$ is compactly supported, i.e. it vanishes outside a finite radius proportional to~$h$.

Derivatives of fields are obtained by applying differential operators to the kernel. For example, a gradient is approximated as follows:

\begin{equation}
\nabla A(r_i) \approx \sum_{j} \frac{m_j}{\rho_j} A_j \cdot \nabla W(r_i - r_j, h_{ij})
\end{equation}

This formulation reduces differential operators to weighted sums over neighbouring particles, with the weights determined by the kernel function.

The choice of kernel strongly influences accuracy and stability. Common examples include the cubic-spline and Wendland kernels, each offering a trade-off between smoothness, computational cost, and neighbour count. In practice, the number of neighbours per particle is kept approximately constant across resolutions by adapting $h$ to the local density.

The dynamics of compressible fluids are governed by the Navier–Stokes equations in Lagrangian form:

\begin{equation}
\frac{d\rho}{dt} = -\rho \nabla \cdot v
\end{equation}

\begin{equation}
\frac{dv}{dt} = -\frac{\nabla \rho}{\rho} + \text{f}^\text{visc} + \text{f}^\text{ext}
\end{equation}

\begin{equation}
\frac{du}{dt} = \frac{P}{\rho^2} \frac{d\rho}{dt} + \Phi
\end{equation}

where $P$, $\rho$, $v$, and $u$ denote the pressure, density, velocity, and specific internal energy, while $\text{f}^\text{visc}$, $\text{f}^\text{ext}$, and $\Phi$ denote viscous forces, external acceleration, and viscous heating and shock terms, respectively.

In SPH, these continuous equations are discretised into particle interactions. For example, the density evolution can be expressed as:

\begin{equation}
\rho_i = \sum_{j} m_j W(r_i - r_j, h_{ij})
\end{equation}

which, for every particle $i$, accumulates the contributions of all neighbours $j$. Likewise, the momentum equation may be written as:

\begin{equation}
\frac{dv_i}{dt} = -\sum_j m_j \left(\frac{P_i}{\rho_i^2} + \frac{P_j}{\rho_j^2}\right) \cdot \nabla W(r_i - r_j, h_{ij}) + \text{f}_i^\text{visc} + \text{f}_i^\text{ext}
\end{equation}

These discrete equations define the computational workload of an SPH simulation. Each particle requires evaluation of sums over its neighbours, typically numbering between 50 and 200 in three-dimensional problems, depending on the smoothing kernel. This leads to two broad classes of operations: per-particle updates that involve only local state, and pairwise interactions. The latter dominate the cost of a timestep.

\subsection{Implementation}

Designing an efficient SPH simulation is an exercise in engineering trade-offs. Logically, each SPH particle typically carries position, velocity, thermodynamic variables, smoothing length, and auxiliary quantities (e.g., density, pressure, sound speed, acceleration, and timestep size)~\cite{Price:2012:SPH,Lind:2020}. This naturally leads to an array-of-structs layout, which is both intuitive to developers and optimal for selected algorithmic phases such as sorting, reordering, or inter-node communication~\cite{Homann:2018:SoAx,Weinzierl:2015:pidt}. Meanwhile, a struct-of-arrays layout may be preferred for computationally intensive tasks that benefit from vectorisation, efficient cache usage, or coalesced accesses—such as the quadratic kernels, especially when deployed on an accelerator~\cite{Homann:2018:SoAx,Strzodka:2011:AbstractionSoA,Sung:2012:DataLayoutTransformations}. A hybrid approach may also be useful, where the most frequently accessed particle properties are stored *in situ* while less frequently used data is stored in auxiliary structures accessed via particle-held pointers~\cite{Hundt:2006:StructureLayoutOptimisation,Jubertie:2018:DataLayoutAbstractionLayers}.

Data structures that accelerate spatial searches, such as bounding-volume hierarchies or octrees, may either store local particle buffers to aid cache utilisation and vectorisation, or store particle indices or pointers into global buffers to accelerate sorting and minimise memory footprint~\cite{White:1994:FMM,Weinzierl:2015:pidt}. Finally, neighbour search and particle interactions may be implemented using tight distance checks on some platforms—notably CPUs—or using more relaxed neighbourhood–neighbourhood distance checks and interaction masking on accelerators~\cite{Wille:2023:GPU,Nasar:2024:GPU}. These choices are not necessarily orthogonal, either in code structure or performance impact, and are often fixed early in a project, becoming increasingly difficult and error-prone to modify as the codebase grows in complexity~\cite{Fowler:2019:Refactoring}.

Our simulation model assumes a conservative baseline typical of modern SPH codes. We work with either contiguous or scattered AoS particle data and permit particle properties to be accessible either directly or indirectly via particle-held pointers. We assume that algorithmic phases are encoded using local or pairwise functions that accept particle references and modify their state directly, and that the control flow immediately surrounding kernel invocations is governed by (possibly nested) \texttt{for} or range-based \texttt{for} loops. Table~\ref{table:particle} lists per-particle quantities and their storage formats.

\begin{table}[htb]
\caption{Particle data format \label{table:particle}}
\centering
\begin{tabular}{lll}
\hline
\textbf{Quantity} & \textbf{Format} & \textbf{Description} \\
\hline
$x$        & float[3] & Position \\
$v$        & float[3] & Velocity \\
$u$        & float    & Internal energy \\
$m$        & float    & Mass \\
$h$        & float    & Smoothing length \\
$\rho$     & float    & Density \\
$P$        & float    & Pressure \\
$c_s$      & float    & Sound speed \\
$a$        & float[3] & Acceleration \\
$\Delta u$ & float    & Thermal advection \\
$\Delta t$ & float    & Timestep \\
\hline
\end{tabular}
\end{table}

\begin{algorithm}[htb]
\caption{Linear and quadratic kernel invocation \label{algorithm:vanilla-kernels}}
\begin{algorithmic}[1]
\Procedure{CallLinear}{$\text{Kernel}$,\ ${i}$}
\ForAll{$p_i \in {i}$}
\State $\text{Kernel}(p_i)$
\EndFor
\EndProcedure

\Procedure{CallQuadratic}{$\text{Kernel}$,\ ${i}$, ${j}$}
\ForAll{$p_i \in {i}$}
\ForAll{$p_j \in {j}$}
\State $\text{Kernel}(p_i,\ p_j)$
\EndFor
\EndFor
\EndProcedure
\end{algorithmic}
\end{algorithm}

\replaced[id=R5]{Table~\ref{table:particle} lists}{Algorithm~\ref{alg:vanilla} demonstrates} both the linear and quadratic kernel-invocation procedures. Within this model, identifying offload points is straightforward: when a single \texttt{Kernel} invocation is too small to amortise driver-launch and interconnect costs, the surrounding loop nest becomes the natural unit of offload. Achieving high performance, however, hinges on kernel- and architecture-specific decisions. First, decide whether to materialise data views—buffer copies that contain only the fields required by the kernel. Next, choose the target layout (AoS or SoA) and where the conversion occurs: a host-side pack that minimises traffic by sending only the required fields, or a device-side transform that trades larger transfers for parallel conversion on the accelerator. Finally, account for interconnect bandwidth and device-memory limits by considering reduced-precision encodings for stored and transported data.

We address this challenge at the language level by designing a set of C++ features that allow developers to decorate loops to specify if and how they should be deployed onto an accelerator.

%% file: 03_cpp_extensions.tex
C++11 \replaced[id=Ours]{introduces the}{introduced the standard} attribute syntax \verb|[[...]]|, which allows declarations or statements to be annotated with additional metadata. Compilers are required to ignore attributes they do not recognise. 
Name\-spaced attributes (e.g. \verb|[[clang::]]|) are \replaced[id=Ours]{compiler implementation-specific}{vendor-specific}; \deleted[id=Ours]{other} compilers \replaced[id=Ours]{that do not support them should}{ typically accept and} ignore them, emitting at most a warning. This makes attributes a portable carrier of optional semantics without breaking compilation on unsupported toolchains.

\subsection{General design}

The design of the novel attributes follows our prior work~\cite{Radtke:2024:AoS2SoA,RadtkeWeinzierl:CCPEAoSSoA,Radtke:2027:CPP}: developers \deleted[id=Ours]{can} maintain their AoS-based codes and simply annotate \texttt{for} and \texttt{for-range} loops to request data-layout transformation, stripping, and computation offloading. 

\added[id=Ours]{These annotations trigger temporary, local, and automatic rewrites through} \verb|[[clang::soa_conversion]]| \added[id=Ours]{and} \verb|[[clang::soa_conversion_compute_offload]]| \added[id=Ours]{implemented by the compiler. 
As in~\cite{RadtkeWeinzierl:CCPEAoSSoA}, we do not assume specific performance characteristics of target architectures or domain-specific precision requirements (e.g.~by picking grain sizes). 
Instead, we treat these attributes as tools for rapid prototyping and validation of high-level, error-prone optimisations across diverse codebases and hardware architectures.
}

\added[id=Ours]{
 Besides the loop transformations, there are also annotations to attributes of structs:
}
\verb|[[clang::truncate_mantissa(N)]] double myattr;|
\added[id=Ours]{
 tells the compiler to hold \texttt{myattr} with a reduced precision with \texttt{N} mantissa bits in memory rather than as a built-in language type.
}
\added[id=R1]{
 This reduced precision format is for storage only. 
 When data is loaded from a buffer into registers for computation, it is automatically expanded to a form supported natively by the hardware, such that all floating-point operations are hardware-accelerated. 
 Once the computation finished and the result is saved back to the buffer/variable, the data is truncated according to the user-specified argument \texttt{N}.
 Within the struct, the compiler eliminates fill-in (padding) bits and bytes to reduce the memory footprint of the host class.
 If we trigger an SoA conversion, the stream over \texttt{myattr} instances becomes a long bitstream, where every \texttt{N} bits encode one value.
 We convert just in time.
}

\deleted[id=Ours]{Specifically, we introduce two new attributes that complement the previously proposed set:}
The attribute \verb|[[clang::soa_conversion_handler]]| may be used in conjunction with \verb|[[clang::soa_conversion_compute_offload]]| to annotate a \texttt{for} or for-range loop over a C++ container, instructing the compiler to construct an SoA view on the \verb|host| (CPU) before shipping it to the accelerator.
\replaced[id=Ours]{Without it, the compiler keeps the data on the host and relies on the hardware to move it}{, or to ship the raw data}  to the \verb|device| (GPU)\replaced[id=Ours]{. The data remains logically in-place.}{ and convert it to an SoA view there}.

\deleted[id=Ours]{These annotations influence how the temporary, local, and automatic rewrites triggered by \ldots and \ldots are implemented by the compiler. As in~\cite{RadtkeWeinzierl:CCPEAoSSoA}, we do not assume specific performance characteristics of target architectures or domain-specific precision requirements. Instead, we treat these attributes as tools for rapid prototyping and validation of high-level, error-prone optimisations across diverse codebases and hardware architectures.}

\deleted[id=Ours]{To illustrate their usage, consider Algorithm~\ref{algorithm:vanilla-kernels}, now equipped with attributes requesting a 16-bit reduced-precision SoA transformation performed on the CPU, with the resulting kernels executed on the GPU.}

\begin{algorithm}[htb]
\caption{Linear and quadratic kernel invocation amended with our proposed C++ attributes.}
\begin{algorithmic}[1]
\Procedure{CallLinear}{$\text{Kernel}$,\ $\{ i \}$} 
  \newline
  $\verb|[[..._compute_offload]]|$
    \newline
  $\verb|[[..._handler(host)]]|$
  \ForAll{$p_i \in \{i\}$}
    \State $\text{Kernel}(p_i)$
  \EndFor
\EndProcedure
\newline
\Procedure{CallQuadratic}{$\text{Kernel}$,\ $\{ i \}$, $\{ j \}$}
  \newline
  $\verb|[[..._compute_offload]]|$
    \newline
  $\verb|[[..._handler(host)]]|$
  \ForAll{$p_i \in \{i\}$}
  \newline
  $\verb|[[..._hoist(1)]]|$
    \newline
  $\verb|[[..._handler(host)]]|$
    \ForAll{$p_j \in \{j\}$}
      \State $\text{Kernel}(p_i,\ p_j)$
    \EndFor
  \EndFor
\EndProcedure
\end{algorithmic}
\end{algorithm}

In the case of both the linear and quadratic kernels\added[id=Ours]{ (Algorithm~\ref{algorithm:vanilla-kernels})}, the annotations play analogous roles. 
The \verb|[[clang::soa_conversion_compute_offload]]| compound annotation requests the AoS–SoA transformation and compute offload to an accelerator. 
\replaced[id=Ours]{An}{The} auxiliary \verb|[[clang::soa_conversion_handler(host)]]| annotation informs the compiler that the data gathering, reduced-precision conversion, and layout change should all be carried out in host memory, and that only the final result should be shipped over the interconnect to the accelerator. 
Finally, the \verb|[[clang::soa_conversion_hoist(1)]]| annotation ensures that the conversions associated with the inner loop are hoisted (taken out of) the nested region and carried out alongside the transformations requested for the parent loop.

\subsection{Reduced precision SoA}

\deleted[id=Ours]{The reduced-precision SoA stores one “anchor” particle per buffer in full precision, while all remaining floating-point data are stored as reduced-precision deltas to the anchor. Conceptually, for each floating-point particle quantity $q$ (e.g., position, velocity, etc.), a per-buffer anchor value is computed from the AoS buffer, $q_A = A({i})$, where $A$ may compute an average over the buffer, pick the value from a random particle, or return something else. Individual values are computed as $\Delta q_i = Q(q_i - q_A)$, where $Q$ is a lossy quantiser—here, a mantissa truncation. Reconstruction of $q$ is the reverse process: $q_i \approx  \Delta q_i + q_A$, and $q_i$ may be used as is.}

\deleted[id=Ours]{The rationale is that many SPH fields vary smoothly within a local neighbourhood, so differences $q_A - q_i$ have a smaller dynamic range than $q_i$ themselves. Consequently, at a given reduced precision level, storing deltas may preserve more precision than storing $q_i$ directly, lowering device memory footprint and transfer volume while preserving an SoA layout for compute. The added overhead of reconstructing $q$ cannot be discarded at face value, especially if reconstructions of a given value $q_i$ happen multiple times, e.g., for every particle–particle interaction. However, as long as SoA buffers are kept small, and given sufficient registers, it is not unreasonable to assume that anchor values will be kept in registers or a local cache and therefore the reconstruction process will induce at most negligible pressure on the main memory subsystem.}

\deleted[id=Ours]{Similar ideas—using locally smooth behaviour to reduce dynamic range and store quantised deltas relative to a local predictor—have been explored in other HPC contexts \cite{Eckhardt:2016:HierarchicalCompression, Weinzierl:2018:HierarchicalRepresentation}. Our approach adapts this principle to SPH particle data and SoA layouts.}

\deleted[id=Ours]{Both the anchor policy $A$ and quantiser $Q$ can be static or adaptive. In an adaptive scheme, $A$ could compute the dynamic range and then $Q$ could decide on the truncation level to keep the error bounded (or zero). We decide to keep the truncation aggressiveness user-tunable. We acknowledge that managing such dynamically sized buffers is possible, but investigating the management overheads it introduces is best reserved for future work.}

\added[id=Ours] {
Our reduced-precision formats supplement non-IEEE value representations \cite{Radtke:2027:CPP}:
Each value annotated with \texttt{truncate\_mantissa(N)} is stored as a bit-truncated floating-point number, where the sign–exponent–mantissa layout does not strictly follow the IEEE standard.
Instead, we approximate different IEEE storage layouts by taking the smallest suitable IEEE format and truncating it to fit the desired total size (Table~\ref{table:fp_formats}).
Multiple such quantities per AoS instance—either as part of one vector or as different attributes of a struct—are automatically packed into one large bitfield by the compiler.
This reduces the memory footprint per struct instance significantly.
}

\added[id=R4]{
A reduced-precision SoA buffer takes different bit sequences of the attribute of interest and concatenates them.
We end up with one long bit sequence, where individual attributes' starting points might not be aligned with native or even byte boundaries.
However, the resulting bitstream has a minimal footprint, addressing constraints imposed by memory bandwidth.
}

\begin{table}[htb]
 \caption{
  Sign-exponent-mantissa bit structure of delta values for example \texttt{[[..truncate\_mantissa(N)]]} argument values. Where the total size corresponds to a known IEEE layout (64, 32, 16) that format is used. Non-standard sizes approximate the IEEE layouts.
  \label{table:fp_formats}
 }
 \centering
 \setlength{\tabcolsep}{4pt}
 \footnotesize
 \begin{tabular}{lcccccc}
 \hline
 \textbf{Item} & \textbf{64} & \textbf{63..33} & \textbf{32} & \textbf{31..17} & \textbf{16} & \textbf{15..} \\
 \hline
 Sign        & 1  & 1  & 1  & 1  & 1 & 1 \\
 Exponent    & 11 & 11 & 8  & 8  & 5  & 5 \\
 Mantissa    & 52 & 51..24 & 23 & 22..11 & 10 & 9.. \\
 \hline
 IEEE layout & \texttt{double} & - & \texttt{single} & - & \texttt{half} & - \\
 \hline
 \end{tabular}
\end{table}

%% file: 03_formalisms.tex
\subsection{GPU offloading}

Let $\mathbb{P}_{CPU,AoS}^{Com,All}$ denote the baseline or ground-truth data structure:
It represents a sequence of particles.
They are stored as an array of structs on the \replaced[id=Ours]{CPU}{GPU}; the particles comprise all attributes that we model in our programming languages, and compressed floating-point values are stored in a memory-efficient, compressed floating-point format.

Let $f$ denote the baseline compute kernel, which takes a set of SPH particles and updates them.
Formally, we have $f: \mathbb{P}_{CPU,AoS}^{Com,All} \mapsto \mathbb{P}_{CPU,AoS}^{Com,All}$.
This baseline (vanilla) version runs on the CPU and operates over the particles stored as AoS.
It is our ground truth in the sense that it operates on the original data structure and yields the baseline performance.
As it operates on compressed data, every single floating-point access converts the compressed floating-point value into a native IEEE format, performs the calculations, and immediately converts back.

Our work introduces a set of data transfer operators:

\begin{itemize}
\item We define $N$ as a narrowing operator that picks only the attributes read by $f$, i.e.\ yields an output with exclusively those variables. The counterpart $N^T$ is widening.
\item The operator $C_{AoS}^{SoA}$ takes the input data in AoS and writes a new output data structure as SoA. The counterpart is $C_{SoA}^{AoS} = \left( C_{AoS}^{SoA} \right)^T$.
\item The operator $U$ unpacks the compressed floating-point values into their corresponding native format. The inverse $U^T$ takes all native data and transforms them to the reduced-precision format.
\item The operator $M_{CPU}^{GPU}$ moves data from the CPU to the GPU. The counterpart $M_{GPU}^{CPU} = \left( M_{CPU}^{GPU} \right)^T$ fetches data back to the host.
\end{itemize}

\noindent
With the four operators \deleted[id=Ours]{defined over $\mathbb{P}*{CPU,AoS}^{Com,All}$, we obtain $2^4 = 16$ different variants of our data structure.
For example, $C*{AoS}^{SoA}: \mathbb{P}*{...,AoS}^{...,...} \mapsto \mathbb{P}*{...,SoA}^{...,...}$.
Along the same lines}, there are multiple variants of the compute kernel\added[id=Ours]{, although not all are relevant in practice or studied here}.
Each acts over a different input data format, but all derive from the same struct declaration.
The source code of $f$ over narrowed data equals exactly the source code of the baseline.
Likewise, all source-code rewrites for the CPU are exactly the same as their GPU counterparts, yet translated for a different target.

\subsection{CPU kernel variations}

On the CPU, the formalism allows us to construct

\begin{eqnarray}
   f(a) & \approx & 
     (\underbrace{N^T \circ U^T}_{} \circ f \circ \underbrace{U \circ N}_{})(a)    
     \label{equation:variant:cpu_uncompress_explicitly} \\
       & = & 
     (\underbrace{C_{SoA}^{AoS} \circ N^T \circ U^T}_{} \circ f \circ \underbrace{U \circ N \circ C_{AoS}^{SoA}}_{})(a)    
     \label{equation:variant:cpu_soa_conversion} 
\end{eqnarray}

\noindent
where $f$ is a generic symbol, automatically adapting to its input data.
Further variants are possible, but these three are, intuitively, the most promising candidates.

Applying the narrowing $N$ alone makes limited sense.
Therefore, it always appears in combination with another operator in the equations above.
The operators bundled together (underbraces) are typically implemented within one compute kernel by the compiler.

By construction, $f(a)$ alone on the left-hand side of \eqref{equation:variant:cpu_uncompress_explicitly} works directly on compressed floating-point data and arrays of structs.
The equivalent right-hand side realises the kernel by adding a preamble and an epilogue to the actual compute kernel:
The preamble converts the compressed data into native IEEE floating-point entries.
This is only done for attributes from the struct that are read within $f$.
The kernel $f$ itself now works exclusively on native data, while the output is converted back into a compressed format by the epilogue once $f$ has terminated over all structs passed into the kernel.

A third variant in \eqref{equation:variant:cpu_soa_conversion} also converts all required data into a native format prior to the actual kernel call, but it additionally reorders the data entries into a struct of arrays.
The kernel itself consequently has to be rewritten by the compiler, with the hope that the SoA facilitates higher vectorisation than the vanilla $f$ blueprint.

\subsection{GPU kernels}

There are two \deleted[id=Ours]{fundamental} paradigms to make \eqref{equation:variant:cpu_uncompress_explicitly} and \eqref{equation:variant:cpu_soa_conversion} exploit a GPU:
We can either transfer all data first to the GPU and convert there, or we can convert on the host prior to the data transfer.

A conversion on the host seems promising, as

\begin{align}
   f(a) &= (M_{GPU}^{CPU} \circ f \circ M_{CPU}^{GPU})(a)
     \label{equation:variant:gpu_native} \\
     & \approx (M_{GPU}^{CPU} \circ \underbrace{N^T \circ U^T}_{} \circ f \circ \underbrace{U \circ N}_{} \circ M_{CPU}^{GPU})(a)    
     \label{equation:variant:gpu_uncompress_explicitly} \\
     &= (M_{GPU}^{CPU} \circ \underbrace{C_{SoA}^{AoS} \circ N^T \circ U^T}_{} \circ f \circ 
     \label{equation:variant:gpu_soa_conversion} 
     \\
     &  \underbrace{U \circ N \circ C_{AoS}^{SoA}}_{} \circ M_{CPU}^{GPU})(a)    
     \nonumber
\end{align}

\noindent
All three variants \eqref{equation:variant:gpu_native}--\eqref{equation:variant:gpu_soa_conversion} perform well if and only if the input data
$\mathbb{P}_{CPU,AoS}^{Com,All}$ are already stored in contiguous chunks of memory.
Otherwise, the move operator has to gather/scatter manually.

Two further implementations can be constructed by permuting $M_{GPU}^{CPU}$ with $ \underbrace{N^T \circ U^T}$ and its counterpart in \eqref{equation:variant:gpu_uncompress_explicitly} or \eqref{equation:variant:gpu_soa_conversion}:

\begin{align}
   f(a) & \approx
     (\underbrace{N^T \circ U^T}_{} \circ M_{GPU}^{CPU} \circ f  \circ M_{CPU}^{GPU} \circ \underbrace{U \circ N}_{})(a)    
     \label{equation:variant:gpu_uncompress_explicitly_cpu_datamanagement} \\
       &= 
     (\underbrace{C_{SoA}^{AoS} \circ N^T \circ U^T}_{} \circ M_{GPU}^{CPU} \circ f \nonumber \\& \circ M_{CPU}^{GPU} \circ \underbrace{U \circ N \circ C_{AoS}^{SoA}}_{})(a)    
     \label{equation:variant:gpu_soa_conversion_cpu_datamanagement} 
\end{align}

\noindent
In this case, we uncompress, or uncompress and convert into SoA, before we ship data to the accelerator.
All data structure management resides on the host, and the GPU becomes a pure compute service.

\subsection{Hybrid constructions\added[id=Ours]{ and in-situ conversion}}

It is an interesting \added[id=Ours]{future} research question to which degree a hybrid realisation of \eqref{equation:variant:gpu_soa_conversion} and \eqref{equation:variant:gpu_soa_conversion_cpu_datamanagement} can yield superior performance.
If we orchestrate

\begin{eqnarray}
   f(a) & \approx & 
     (\underbrace{N^T  \circ M_{GPU}^{CPU} }_{} \circ \underbrace{C_{SoA}^{AoS} \circ U^T}_{} \circ f 
     \nonumber \\
     && \circ \underbrace{U  \circ C_{AoS}^{SoA}}_{} \circ  \underbrace{M_{CPU}^{GPU} \circ N}_{})(a),
     \label{equation:variant:hybrid_soa_conversion} 
\end{eqnarray}

\noindent
we take the input data and transfer only those particle attributes to the GPU that are needed there.
However, this CPU–GPU data exchange format remains AoS.
Arriving as AoS on the accelerator, the accelerator then uncompresses the data and transfers it into SoA.
\deleted[id=Ours]{Again, we can construct variations of this scheme by, for example, skipping the conversion and issuing only $U$ or $U^T$.}

The appeal of \eqref{equation:variant:hybrid_soa_conversion} is that we reduce the CPU–GPU data exchange to its minimum, \deleted[id=Ours]{benefiting even from the compression.
At the same time, data conversion on the CPU must scatter and gather, but it does not permute any data.
We may assume that it therefore streams the conversion through its memory hierarchy, and caches are optimally exploited.}
\added[id=Ours]{
 but employ the GPU for data arrangements.
 In practice, we omit this variant to avoid the rising gather and scatter operations on the host.
}

\added[id=Ours]{
 Instead, our language extension supports \emph{streaming}, where all data reorganisation happens on on the CPU before it is offloaded,
 or we support \emph{in-place} kernels, where the whole (original) AoS CPU data is shipped off and then subject to manipulations on the accelerator.
 We hypothesize that streaming is advantageous if kernels operate on only few attributes of a struct and are invoked only once.
 When multiple kernels are invoked on the accelerator over a struct one after the other, we asume that is pays off to accept the increased offoad footprint of transferring the whole struct and to optimise data arrangements temporarily on the GPU.
}

%% file: 04_compiler_implementation.tex
We modify a popular C/C++ compiler toolchain called Clang/LLVM to handle the novel attributes and produce highly optimised machine code. Within the compilation pipeline, Clang is responsible for parsing C++ source files and producing an LLVM-specific, language-agnostic intermediate representation called LLVM IR, which is subsequently optimised and translated to executable machine code within LLVM. There are multiple ways our attributes may be handled by the compiler toolchain. For simplicity, we chose to implement the handling entirely within Clang and rely on pre-existing optimisation functionality built into LLVM for performant machine code.

\subsection{Data flow}

Within Clang, the \verb|Parser|, \verb|Lexer|, and Semantic Analysis (\verb|Sema|) modules work together to produce an Abstract Syntax Tree (\verb|AST|), which is subsequently traversed by the Code Generation (\verb|CodeGen|) module to produce the LLVM IR. Together, these actions constitute a \verb|FrontendAction|. We implement the handling of our attributes with a new \verb|FrontendAction|.

Similarly to our previous work, the new \verb|FrontendAction| traverses the AST top down and searches for our annotations. When it encounters a convert, it inserts allocation statements for the out-of-place memory allocations, issues the actual data copying from the original AoS buffers, modifies the loop body to adapt it to the new data layout, and eventually adds an epilogue that synchronises the data and frees the temporary buffers.

We need to identify all read and write accesses to struct members subject to the conversion. Without major modifications to LLVM IR, these changes are language dependent. Therefore, it is reasonable to collate all annotation handling exclusively within the \verb|FrontendAction| rather than scatter it among multiple compile stages.

Automatic identification of data manipulated inside loops (data view) requires reading complete code blocks before constructing tailored view representations. C++’s strict typing does not provide logical data type information from narrowing when first encountering annotated loops. This automatic view identification breaks the principle that complete data information must be available before first usage. Therefore, a preparatory pass collects access data before a second pass applies the required modifications. The AST must be analysed before implementing transformations. This two-pass front-end implementation resembles preprocessor macro handling—a separate stage that remains transparent to other front-end pipeline components.

By abandoning the single-read approach and processing code multiple times, translating annotated code to plain OpenMP-augmented source becomes more practical than integrating GPU offloading directly into translation. We use source-to-source translation followed by OpenMP processing within Clang’s native \verb|FrontendAction|. Though logically separate, our \verb|FrontendAction| maps to OpenMP-annotated C++ code through in-memory source replacements, working around Clang’s nearly immutable AST. After initial parsing, we modify the in-memory buffers containing the original translation unit files, using annotations to guide rewrites. The compiler then re-parses these modified buffers and feeds the new AST into the pipeline. This approach reuses the preprocessor, lexer, parser, and semantic analysis across both passes.

As we navigate the source code using the AST, the effect of our annotations is scoped at the translation-unit level: the data transformations cannot propagate into user translation units (object files) and notably fail to propagate into libraries.

\subsection{Data management}

Out-of-place transformations for containers typically require heap data allocations if the containers’ size is unknown at compile time. Heap allocations can become expensive if issued very frequently in high-concurrency codes \cite{Hager:2011:HPCIntro}. 
However, \replaced[id=Ours]{we do not need}{our prologue–epilogue paradigm does not require strictly} dynamic data structures. 
\replaced[id=Ours]{The struct members used by a kernel}{We know the size of the underlying container by the time we hit the prologue, while the struct member set} is a compile-time property. Since the size of the container does not change while we traverse the loop, we work with variable-length arrays (VLAs)~\cite{Cheng:1995:VariableLengthArrays}. This approach avoids memory allocator calls entirely.

\subsection{GPU data handling}

Device memory allocation on critical paths causes significant performance degradation, requiring a memory-management scheme that safely reuses accelerator allocations. Our implementation uses lazy allocation of per-thread, per-transformed-loop buffers that grow based on working set sizes. To minimise synchronization overhead, we schedule all accelerator operations—including memory transfers and kernel executions—asynchronously using standard OpenMP \verb|nowait depend(...)| constructs. Since the arising prologue–loop–epilogue sequence operates synchronously, we cannot introduce observable asynchronicity beyond the transformed loop’s immediate context. This requires at least one host–device synchronization per loop invocation, implemented using OpenMP’s \verb|taskwait| construct. The synchronization combined with per-thread, per-loop buffer allocation ensures data safety.

\subsection{Interplay with other translation passes and shortcomings}

By confining our transformation entirely within Clang, we ensure that all optimisations performed by the LLVM optimiser as part of subsequent passes are applied to the transformed kernels. At this point, our transformations are already out of the way. This guarantees that each kernel is processed in a single, consistent form, eliminating potential conflicts arising from multiple interpretations of the same code: if functions are used within views and then also without views, we have logically already replicated the functions’ realisation, and the optimisations handle the replica independently.

Such a design not only avoids unfortunate side effects; it also ensures that HPC-critical optimisations, such as auto-vectorisation, are applied independently and in a bespoke, tailored way to the vectorisation-friendly SoA representations.

A limitation of our approach is its dependence on complete visibility of all functions that interact with the transformed AoS data. We need access to the whole loop body’s call graph to identify all write and read accesses to AoS data members and to produce a modified loop kernel. Making the whole call hierarchy visible becomes challenging when external functions, either directly or indirectly referenced by the kernel, reside outside the analysed translation unit. In such cases, our transformations break down.

%% file: 05_results.tex
\begin{figure*}[h]
\centering
\includegraphics[width=8cm]{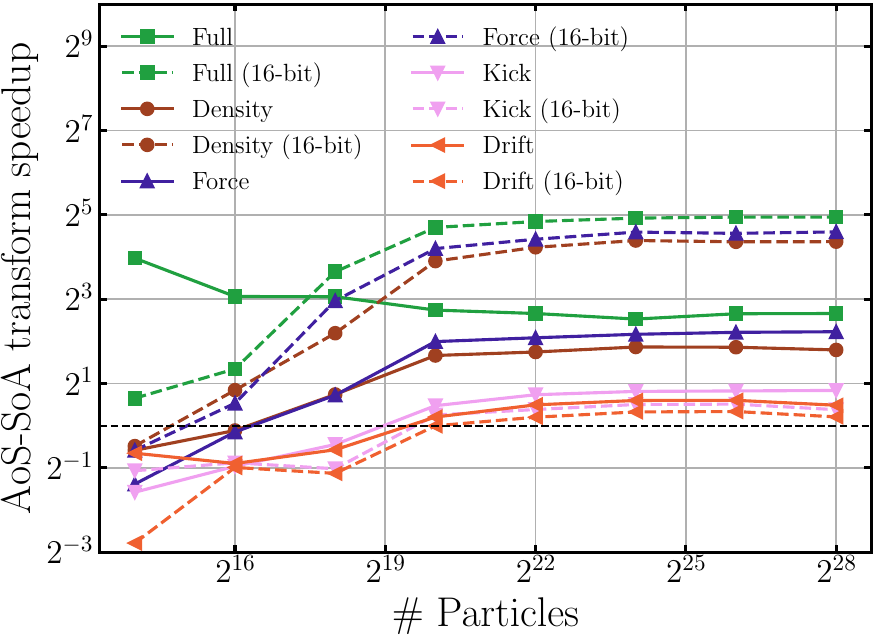}
\includegraphics[width=8cm]{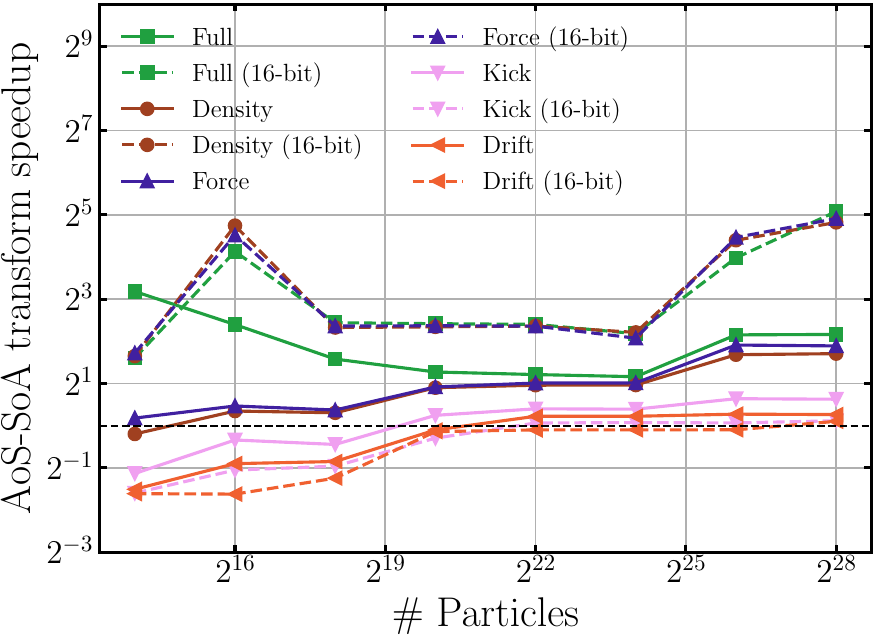}
\includegraphics[width=8cm]{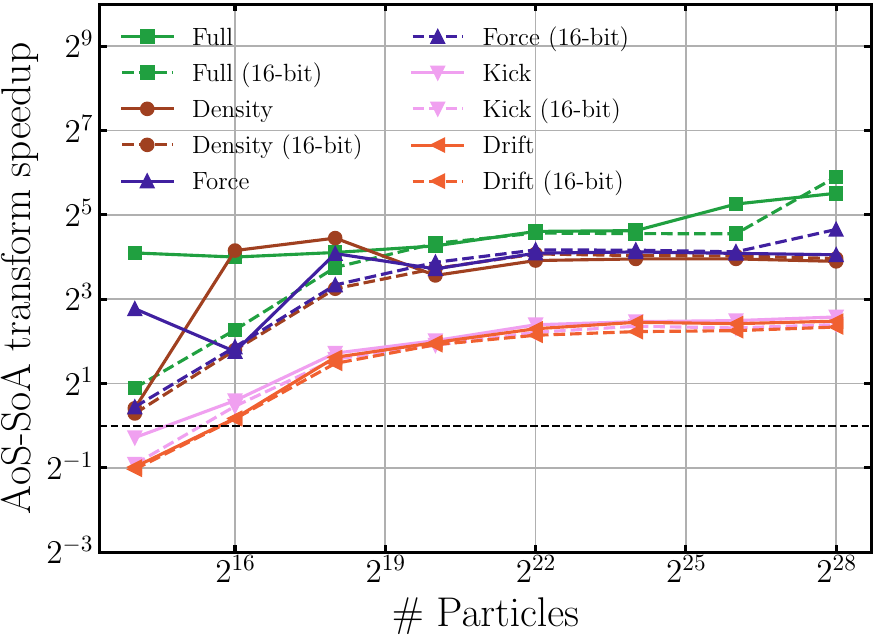}
\includegraphics[width=8cm]{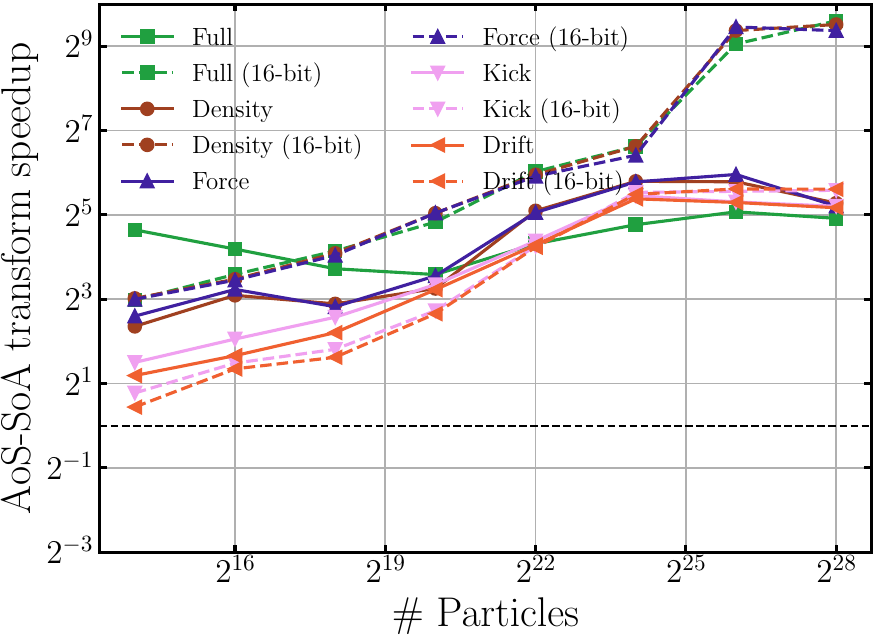}
\caption{AoS–SoA layout transformation speedup on GPU relative to the CPU baseline for the H100 node (top left), the MI200 node (top right), the GH200 node (bottom left), and the MI300A node (bottom right). The 16-bit variants use the reduced-precision SoA buffers\deleted[id=Ours]{, which store floating-point data as 16-bit deltas}. Values above $1$ indicate that the GPU-based conversion is faster, whereas values below $1$ mean the CPU-based conversion is more performant.}
\label{fig:transform}
\end{figure*}

We evaluate the impact of the proposed language extensions across a diverse set of accelerator hardware configurations. The H100 node integrates dual Intel Xeon Gold 6530 CPUs and an Nvidia Hopper H100 GPU. The GH200 node represents an Nvidia Grace Hopper architecture. The MI200 node comprises a dual-socket AMD EPYC 7513 system paired with an AMD Instinct MI200 GPU. Finally, the MI300A node features four AMD MI300A APUs. To ensure consistency and mitigate the influence of non-uniform memory access (NUMA) effects, all tests are confined to a single NUMA domain per system. Furthermore, to prevent resource contention and oversubscription of the accelerators, the number of CPU cores allocated for offloading tasks is standardised at 16 across all platforms.

The particle size in the double-precision baseline is 272 bytes, compared to 144 bytes in the single-precision baseline. This corresponds to round-trip offloading costs of 544 bytes and 288 bytes, respectively. The ratio between these baselines is not precisely 1:2 due to the inclusion of non-floating-point data, such as a 64-bit integer ID and position data, which remain in double precision regardless of configuration.

All particle data are allocated on the host CPU and stored as AoS with compressed floating-point representations. Prior to the invocation of the compute kernels, we allocate memory on the GPU that is large enough to hold all data in a non-compressed format, i.e. large enough to host all structs if they are extended into a realisation relying on double precision everywhere. This memory is allocated once and never released, as dynamic memory management is known to be expensive and, hence, memory pooling is a well-established technique in GPU programming even if data are not GPU-resident \cite{Wille:2023:GPU}. In the absence of managed shared memory, we manually copy data to and from the GPU through the OpenMP copying routines into the pre-allocated buffers. Alternatively, we could have employed \texttt{\#pragma omp data enter} and \texttt{\#pragma omp data exit} OpenMP pragmas with explicit mapping. The GPU kernel invocation is then a simple \texttt{\#pragma omp target teams distribute parallel for}.

We measure performance by timing the execution of SPH kernels adapted from SWIFT \cite{Schaller:2024:Swift}, applied to particle buffers of varying sizes. 
\added[id=R5]{Two quadratic kernels---Density and Force computing the respective quantities density and force---compute neighbourhood-dependent quantities based on pairwise interactions such as density, pressure, acceleration, or thermal advection. 
The linear kernels Kick and Drift locally update quantities such as position, velocity, or internal energy.}

The data are stored in an array-of-structures (AoS) format, where all attributes are placed directly within the particle struct. No auxiliary data linked through particle-held pointers are used. All floating-point quantities are stored in 32-bit precision, except for particle positions, which are maintained in 64-bit precision.

Particles are grouped into contiguous buffers representing spatially co-located neighbourhoods. Each buffer contains 64 particles, matching the typical neighbour count employed in many SPH implementations. Linear kernels such as Kick and Drift process each particle once per time step, while quadratic kernels such as Density and Force perform 64 pairwise interaction calculations for each particle.

\subsection{Performance of AoS–SoA transformation on CPU vs GPU}

We begin by evaluating the performance of AoS to SoA transformations on both CPUs and GPUs. \replaced[id=R1]{We measure the relative performance of the memory layout conversion and data transfer only, leaving the impact assessment on accelerating computation to the next subsection. The CPU data time the beginning of transformation (\ref{equation:variant:gpu_soa_conversion_cpu_datamanagement})---$M_{CPU}^{GPU} \circ U \circ N \circ C_{AoS}^{SoA}$, whereas the GPU data time the beginning of (\ref{equation:variant:gpu_soa_conversion})---$U \circ N \circ C_{AoS}^{SoA} \circ M_{CPU}^{GPU}$}{The measurements cover the entire transformation pipeline, including all data transfers}. The narrowing operator $N$ is tailored to the requirements of a specific kernel, so the resulting SoA buffers contain only the fields necessary for that kernel’s execution. For comparison, we also include a Full variant, which transforms all particle fields irrespective of kernel usage.

Figure \ref{fig:transform} shows the relative performance of the memory layout conversion executed on the GPU vs on the CPU for the four tested architectures. The most pronounced speedups are observed in the Full, Density, and Force variants. This trend aligns with the fact that both Density and Force kernels make use of most particle attributes. As a result, transferring the complete set of particle data to the GPU for transformation does not introduce a disproportionate overhead. In such cases, the cost of moving all fields is amortised by the kernels’ broad data usage, making the GPU-based transformation particularly performant.

In contrast, the Kick and Drift kernels, which rely on only a small subset of particle attributes, see more modest benefits. This effect is especially noticeable on PCIe-based accelerators, where the limited host–device bandwidth amplifies the cost of sending all particle data to the GPU for conversion, even though only a fraction of those fields is actually required. For such kernels, the overhead of unnecessary data transfers eclipses the advantage of performing the transformation on the accelerator.

\replaced[id=R1]{The impact of reduced precision varies across kernel types. The Density and Force kernels benefit the most, and the impact is greater on fast-interconnect architectures than on classic PCIe ones. Since both Density and Force kernels use most particle data, the data efficiency per loaded cache line is reasonably good, and so bandwidth becomes a significant factor. On the other hand, Kick and Drift kernels consume few particle properties, which leads to poor data efficiency per loaded cache line during the conversion, and so the load latency — which is unaffected by reduced precision in this workload — dominates, leading to a lack of performance improvement across all architectures.}{Another clear trend is that reduced-precision variants consistently outperform their full-precision counterparts. This outcome is expected, as the transformation process is dominated by memory traffic. Reducing the volume of data through lower-precision representations directly translates into higher throughput and shorter end-to-end transformation times despite the added arithmetic overhead necessary to calculate the reduced-precision deltas.}

Overall, GPU-based transformations deliver substantial performance improvements compared to CPU execution. The gains are especially visible on accelerators equipped with high-bandwidth interconnects, such as the Nvidia GH200 and AMD MI300A, where we observe speedups of up to 35× for the 32-bit precision variant and up to 500× for the 16-bit reduced-precision variants.

\subsection{Impact of AoS vs SoA on kernel performance}

\begin{figure*}[t]
\centering
\includegraphics[width=8cm]{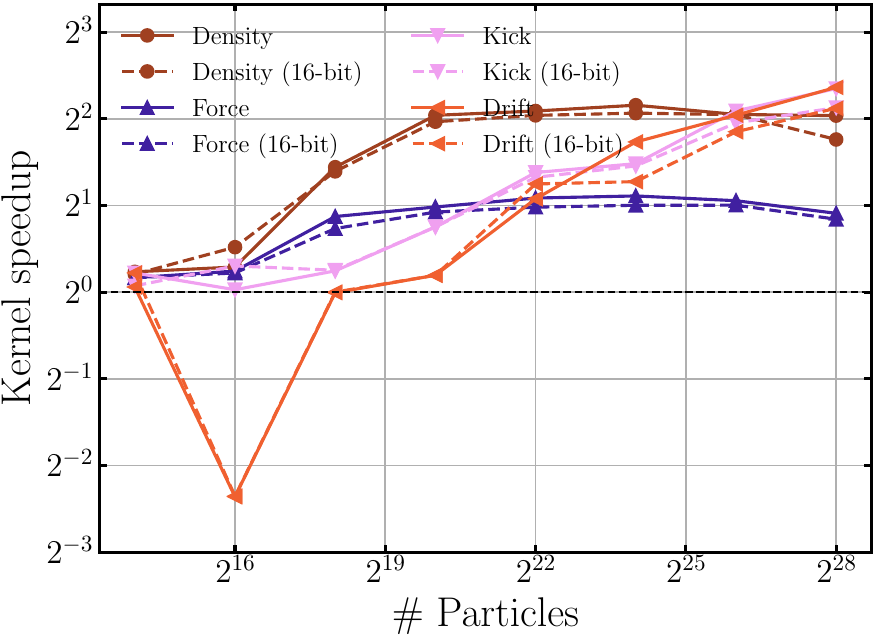}
\includegraphics[width=8cm]{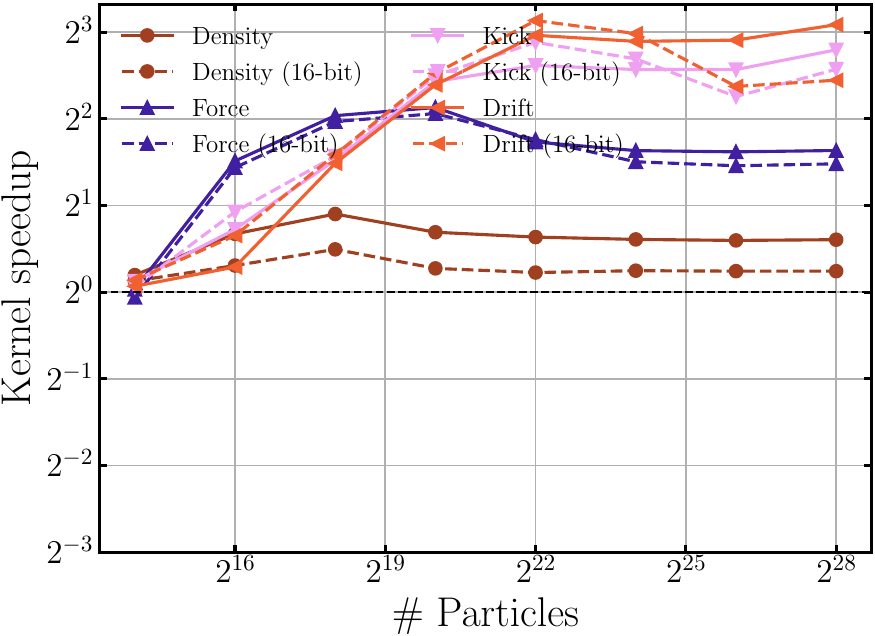}
\includegraphics[width=8cm]{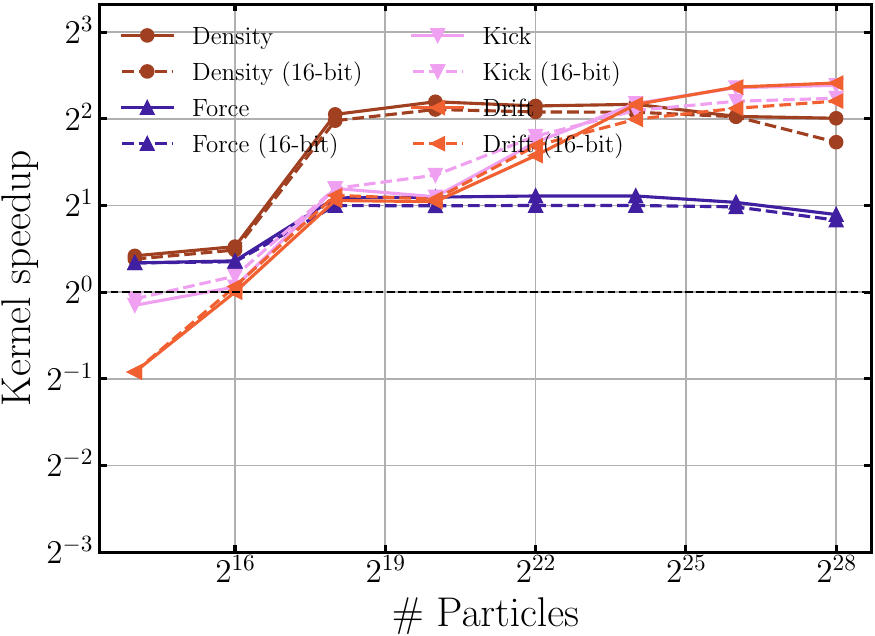}
\includegraphics[width=8cm]{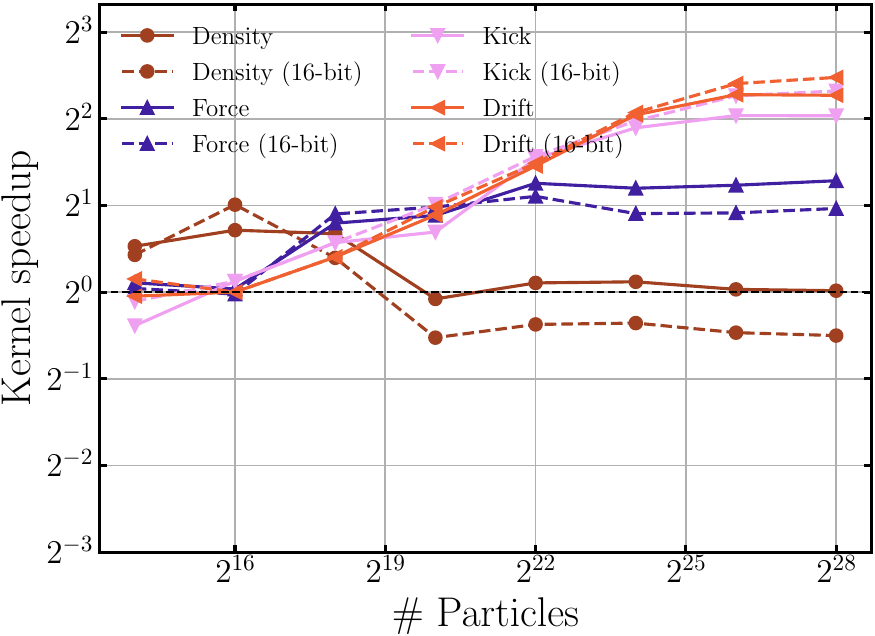}
\caption{SPH kernel compute speedup using SoA vs AoS layout for the H100 node (top left), the MI200 node (top right), the GH200 node (bottom left), and the MI300A node (bottom right). The 16-bit variants use the reduced-precision SoA buffers. Values above $1$ indicate that the SoA layout provides a speedup, whereas values below $1$ mean the vanilla AoS layout is more performant.}
\label{fig:compute}
\end{figure*}

We proceed to evaluate the performance of SPH compute kernels consuming data in array-of-structures (AoS) versus structure-of-arrays (SoA) layouts. These measurements capture solely the execution time of the compute kernels $f$ \added[id=R1]{from (\ref{equation:variant:gpu_soa_conversion})} and omit any data transfers between the host and device. The data show speedups over the baseline where computations consume AoS buffers.

Figure \ref{fig:compute} shows the speedup gained by consuming data in SoA format. In most cases, adopting a SoA layout improves performance, although the degree of speedup depends heavily on the kernel. Kernels dominated by memory operations, such as Kick and Drift, benefit the most from this transformation, as they are highly sensitive to memory coalescing and achieve much higher throughput when data are laid out contiguously by attribute.

The Force kernel, by contrast, shows more modest gains. This result is consistent with its high arithmetic intensity: the kernel is more compute-intensive than memory-intensive, so the benefits of more efficient memory access are less pronounced. The Density kernel sits between these extremes in terms of arithmetic intensity. On Nvidia hardware, its performance scales as expected, yielding a clear improvement. On AMD accelerators, however, Density underperforms relative to the baseline, indicating that hardware-specific memory or compute balance can alter the expected behaviour.

The impact of reduced precision also varies by vendor. On Nvidia systems, reduced-precision variants achieve performance that is broadly consistent across all kernels, indicating that the smaller data volume provides benefits without introducing additional penalties. On AMD hardware, however, the quadratic kernels are negatively affected. The Force kernel suffers a 10–20\% penalty, while Density shows an even stronger degradation of up to 50\%. In this case, the reduced-precision Density kernel performs worse than the vanilla AoS baseline, underscoring that precision reduction is not universally beneficial.

Overall, the largest speedups are observed for the Kick and Drift kernels. On the GH200 platform, performance improves by a factor of eight, while other systems show speedups closer to four. The Force kernel achieves a more moderate improvement, with factors of three on the GH200 and about two elsewhere. The Density kernel exhibits the greatest variability: speedups of roughly four are achieved on Nvidia hardware, but AMD systems show improvements ranging from 1.5× down to a performance regression of about 0.75×.

\subsection{Combined performance impact}

\begin{figure*}[t]
\centering
\includegraphics[width=8cm]{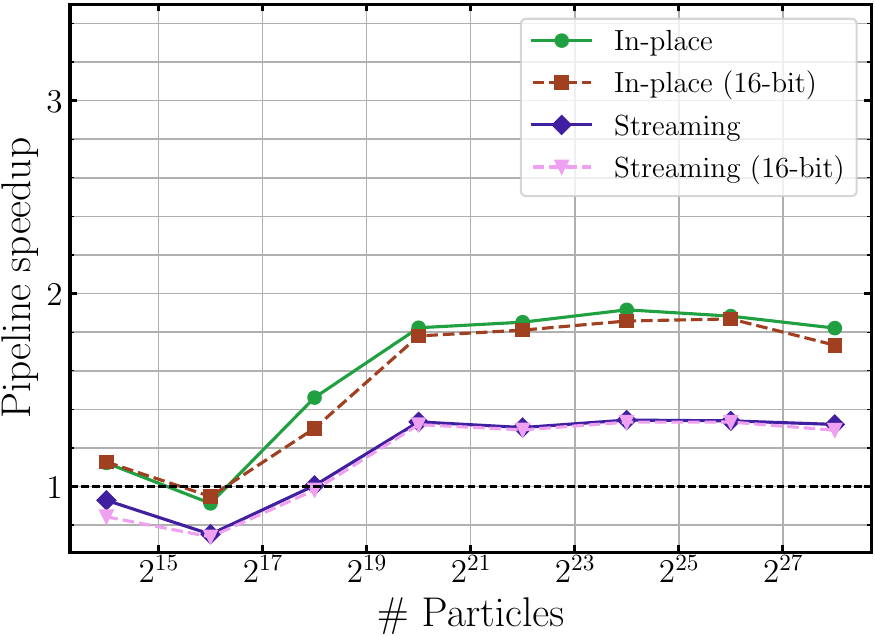}
\includegraphics[width=8cm]{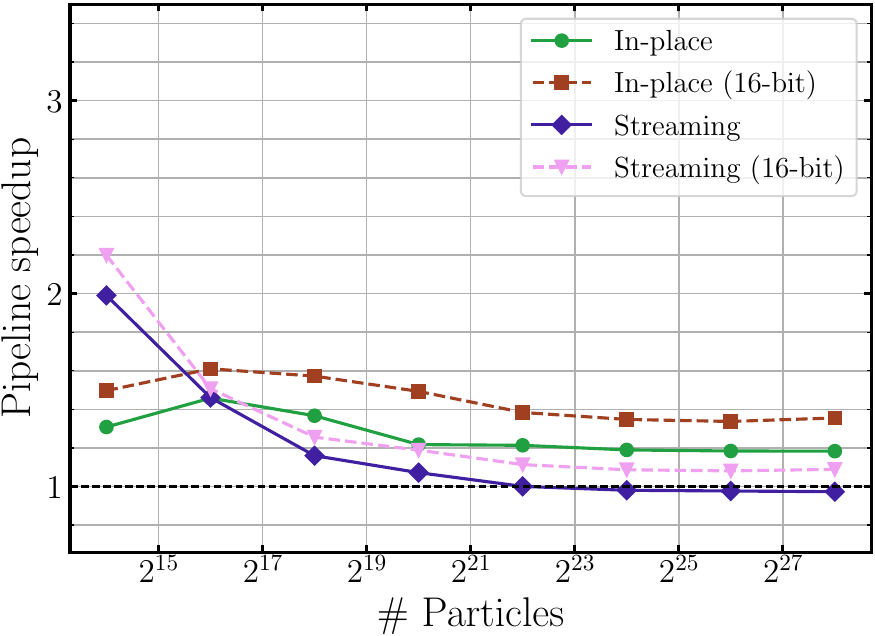}
\includegraphics[width=8cm]{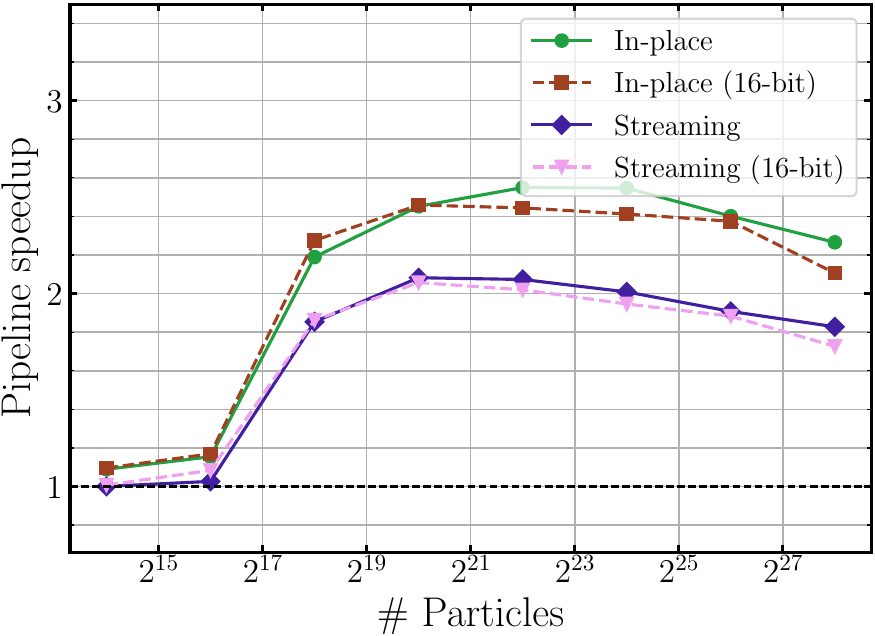}
\includegraphics[width=8cm]{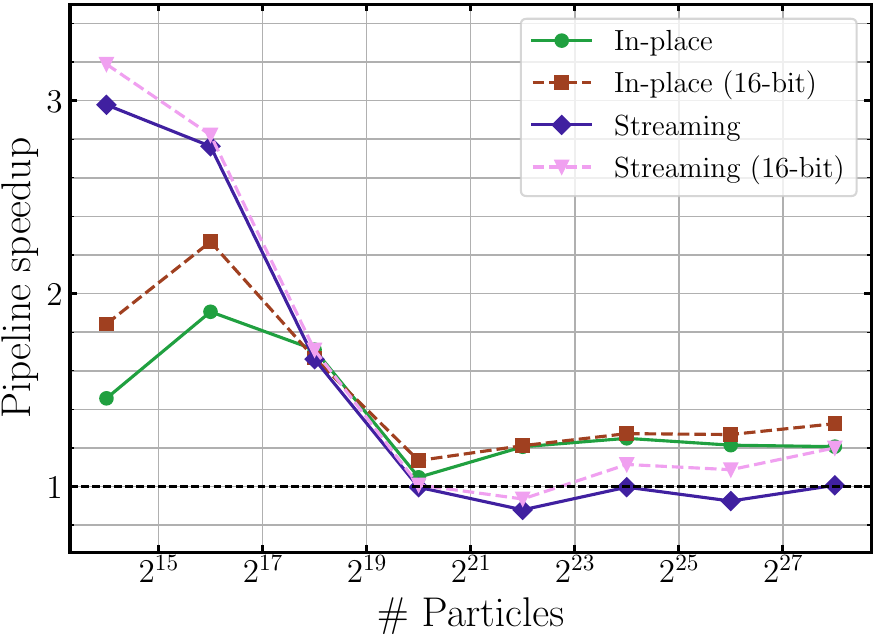}
\caption{End-to-end speedup of GPU-based SoA vs AoS baseline for the H100 node (top left), the MI200 node (top right), the GH200 node (bottom left), and the MI300A node (bottom right). The GPU-based transformation is chosen as the more performant based on the results above. The 16-bit variants use the reduced-precision SoA buffers. Values above $1$ indicate that the AoS–SoA transformation provides an end-to-end speedup, whereas values below $1$ mean that the vanilla computation consuming AoS data is more performant.}
\label{fig:pipeline}
\end{figure*}

We conclude our testing by evaluating the combined effect of AoS–SoA transformation placement, reduced-precision layout, and kernel compute times on the performance of the entire pipeline (time step). We \replaced[id=Ours]{examine two ways how to employ our transformations}{model two simulation architectures}: 
\emph{In-place} assumes that whole particle buffers are transferred to an accelerator\added[id=Ours]{ once}, kept there while all four kernels are invoked, and then shipped back to the host. In contrast, the \emph{Streaming} arrangement assumes that each kernel invocation requires data to be continuously moved between host and device—and furthermore, that only the data needed for a given kernel invocation are sent across.

Figure \ref{fig:pipeline} shows the speedup gained by employing \added[id=R1]{the entire transformation from (\ref{equation:variant:gpu_soa_conversion})}---the GPU-based AoS–SoA transformation---over \replaced[id=Ours]{our}{a non-transforming} AoS baseline for the four tested architectures. \added[id=R3]{Time-step behaviour does not mimic the patterns observed in earlier sections, and the overall runtime improvements are more modest. In part, this is caused by uneven runtime contributions: the Force kernel takes up on average $63.1\%$ of runtime, the Density kernel $36.2\%$, whereas the Kick and Drift kernels together contribute less than $1\%$. Furthermore, even though the pure arithmetic computation is accelerated as per our kernel-level benchmarks, the total cost of the transformation pipeline—including reduced-precision packing ($U$ and $U^T$), narrowing ($N$ and $N^T$), and memory layout change ($C_{SoA}^{AoS}$ and $C_{AoS}^{SoA}$)—incurs non-negligible costs that limit the net performance gains over the vanilla, transformationless baseline.}

A consistent trend across all platforms is that reduced precision has only a limited impact on performance. The only clear exception is the MI200 system, which shows noticeable gains from precision reduction. On all other hardware, the difference between full-precision and reduced-precision execution remains small enough to have little effect on the overall performance.

Beyond this, the influence of reduced precision appears to be more closely tied to the vendor than to the specific hardware architecture. On Nvidia accelerators, all variants outperform the AoS baseline. The magnitude of improvement is greater on the GH200 node than on the H100. On GH200, the In-place variant achieves a speedup of about 2.6×, while the Streaming variant improves by roughly 2×. On the H100, the corresponding gains are 1.9× and 1.4×, respectively, highlighting a consistent advantage for the In-place method.

The AMD platforms show a more counter-intuitive pattern. Reduced-precision variants perform relatively well at small particle counts, but the benefits diminish quickly as the working set grows. Consequently, the observed gains range from negligible to about 1.4× across both AMD systems. Among the tested configurations, the In-place 16-bit variant emerges as the most effective, although the improvements remain modest compared to those observed on Nvidia hardware.

%% file: 08_correctness.tex
\begin{figure*}[htb]
 \begin{center}
  \includegraphics[width=0.3\textwidth]{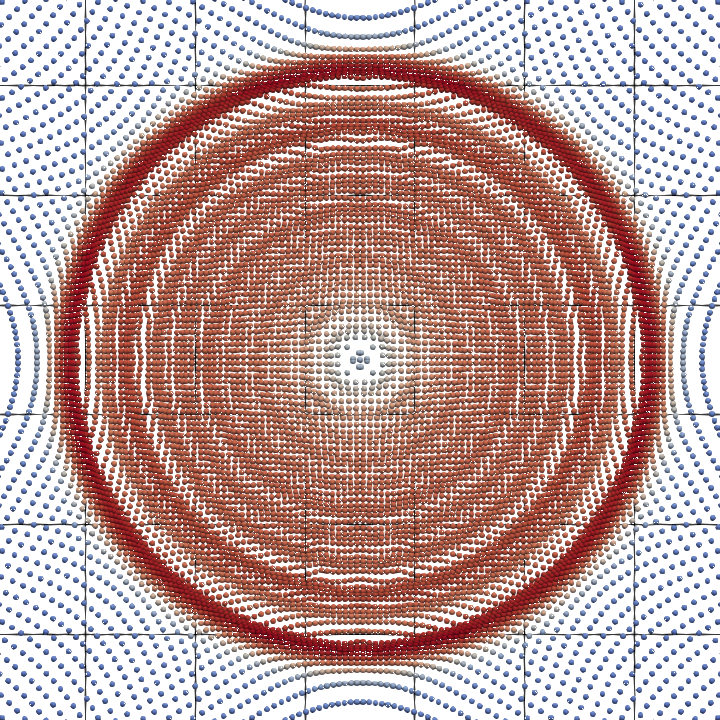}
  \includegraphics[width=0.3\textwidth]{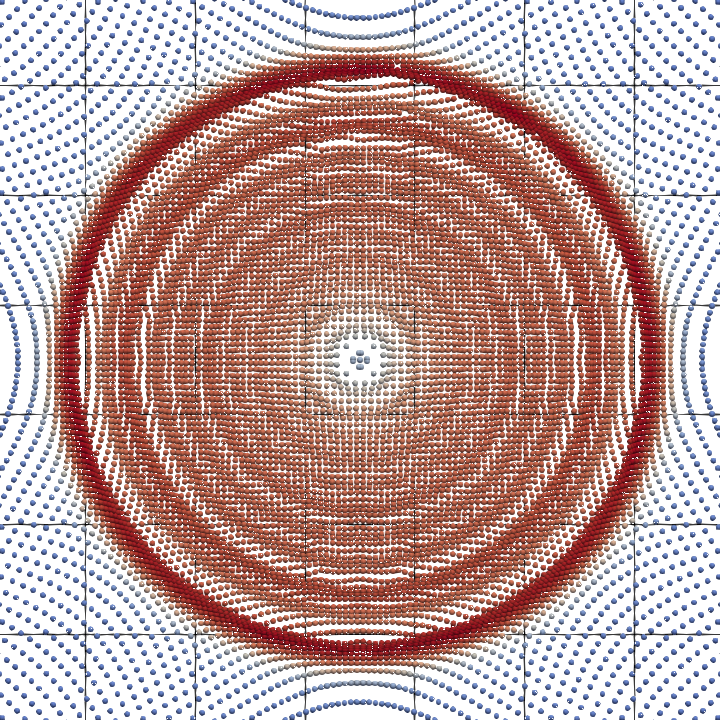}
  \includegraphics[width=0.3\textwidth]{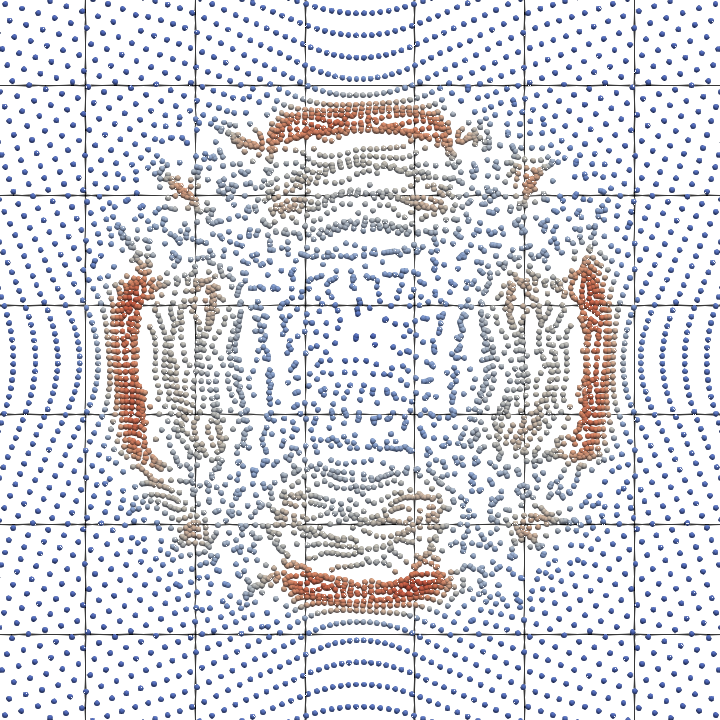}
 \end{center}
 \caption{
  \added[id=R1]{
   Central region of the Noh problem in 2D at $t=0.1$ for simulation runs using full 64-bit storage precision (left), 32-bit reduced-precision (center), and 16-bit reduced precision (right) from our related work\cite{Radtke:2027:CPP}. Truncation below 32-bit precision leads to non-physical symmetry breaking of the experiment.
   \label{figure:correctness:noh2d} 
  } 
 }
\end{figure*}


\begin{figure}[htb]
 \begin{center}
  \includegraphics[width=0.48\textwidth]{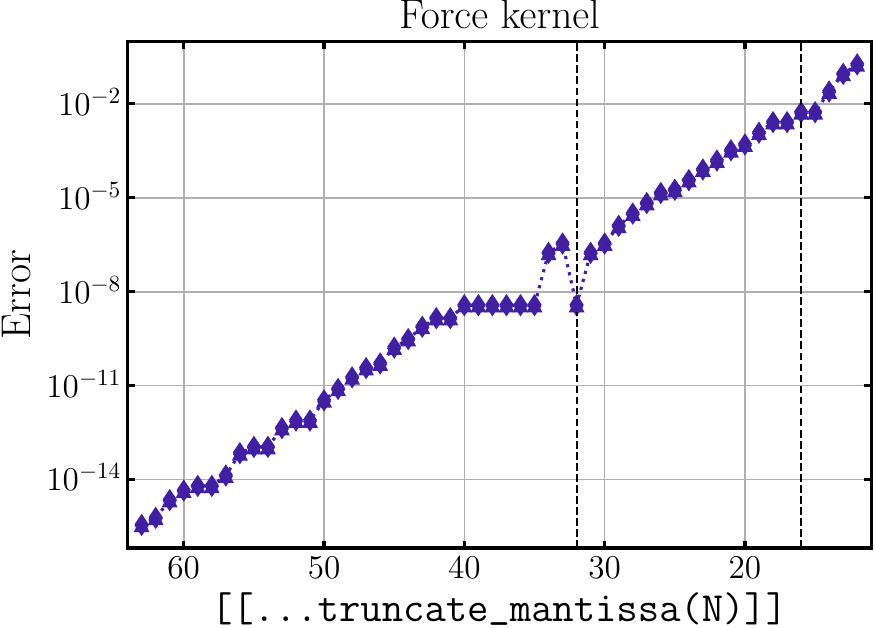}
 \end{center}
 \caption{
  \added[id=R1]{
   Average per-particle acceleration error introduced by a single invocation of the Force kernel over increasingly more compressed floating-point data vs a full 64-bit precision reference. Vertical lines at $N=32$ and $N=16$ mark the transition points to the single and half precision IEEE layouts as the base for truncation, respectively.
  } 
  \label{figure:correctness:force}
 }
\end{figure}

\added[id=R1]{
 We close our discussion with an analysis of the impact of floating-point compression.
 For this, we first run all compute kernels with double precision over approximately 65,000 particles, and
 store the quantities that each kernel alters.
 For the force kernel, this is the resulting acceleration.
 This provides us with a correct baseline, i.e.~a reference solution that carries only the standard IEEE 64-bit truncation error.
}

\added[id=R1]{
 When we truncate the floating-point numbers (Figure~\ref{figure:correctness:force}), the relative root mean square error (RMSE) of the acceleration---i.e., the error divided by the average of the quantity of interest---increases as we reduce the number of valid mantissa bits.
 The data for the density calculations resemble the force results, and all measurements are relatively consistent, i.e., tracking the maximum error instead of the RMSE yields very similar outcomes (not shown).
 As long as we retain more than approximately 50 bits per floating-point value, the resulting error in the acceleration vector remains within the regime of machine precision.
 Since our ground truth baseline is also only accurate up to machine precision, we may conclude that mantissa truncation does not introduce an observable error at this level.
}

\added[id=R1]{
 Down to a precision of around 40 bits, the truncation error increases monotonically as we strip away more bits from the mantissa.
 After that, it hits a plateau.
 The plateau's error is in the range of single precision, before the error curve rejoins the deterioration trend again at 33 and 34 bits.
 The plateau is an outlier from the overall trend.
}
 
\added[id=R1]{
 As we go below 32 bits, we toggle the representation of the exponent (Table~\ref{table:fp_formats}).
 Due to this switch, we obtain relatively high accuracy at exactly 32 bits, where there is no actual truncation in software.
 We match the IEEE format:
 This is a sweet spot where a high number of valid bits relative to the bits invested in the exponent suddenly becomes available.
 After that, the error curve rejoins the previous trends, although the curve is offset—i.e., slightly improved compared to the 33+ bit trend.
}
 
\added[id=R1]{
 The same pattern repeats, albeit less pronounced, around the switch from single to half precision in the exponent representation.
 We may assume that more sophisticated truncation formats, where the exponent does not switch abruptly, would yield overall more accurate results.
 Converting the data into SoA has no impact on the outcomes.
 This is reasonable given that the conversion is solely a memory organization rearrangement without any additional truncation.
}

\added[id=R1]{
 We conclude with a reference to studies from \cite{Radtke:2027:CPP}, where we run the Noh 2D benchmark over a longer time span (Figure~\ref{figure:correctness:noh2d}).
 While full 64-bit precision as well as half precision yield rotationally symmetric solutions, reducing the storage precision further destroys these symmetries.
 Our per-kernel truncation error from Figure~\ref{figure:correctness:force} destroys the solution qualitatively, i.e., the errors reported for individual kernel evaluations introduce long-term pollution.
 Mixed storage choices holding particle positions in double precision but applying truncation to selected other particle properties yield visually more appropriate solutions (cf.~work in \cite{Radtke:2027:CPP}).
 While a quantitative analysis of these effects and studies over appropriate precision choices are beyond our scope here, our work introduces a toolset to streamline such fine-tuning of chosen storage formats.
}

%% file: 09_conclusion.tex
This study examined the performance of AoS-to-SoA transformations and reduced-precision data layouts across a range of particle-based kernels and hardware platforms. The results show that the performance impact is governed by a combination of kernel arithmetic intensity, memory access patterns, and interconnect bandwidth. Kernels dominated by memory traffic, such as Kick and Drift, achieved the highest speedups, whereas the compute-intensive Force kernel showed more modest improvements. The Density kernel, with intermediate arithmetic intensity, exhibited a mixed outcome: it performed as expected on Nvidia hardware but showed reduced or inconsistent gains on AMD platforms.

Reduced precision was found to have only a limited effect overall. The MI200 was the only system to show a consistent benefit, while other platforms demonstrated either negligible or strongly vendor-dependent behaviour. 
\added[id=Ours]{
Modern GPUs are well-equipped with fast memory and, different to CPUs with their deep caches, hence do not benefit that significantly from memory footprint reductions on the device.
}
On Nvidia hardware, both reduced-precision and SoA variants outperformed the AoS baseline, with the GH200 node achieving the strongest results. In particular, the In-place transformation strategy delivered the most consistent gains, reaching up to 2.6× on GH200, compared to 1.9× on H100. On AMD systems, however, performance gains from reduced precision were limited to small particle counts, with improvements vanishing as the working set grew. The best case on these platforms was a modest 1.4× speedup for the In-place 16-bit variant.

Future work may focus on developing adaptive transformation strategies that take into account kernel characteristics and hardware profiles at runtime. This includes exploring hybrid precision schemes that selectively reduce data volumes based expected error sensitivity, available cache volumes, and the dynamic balance between compute throughput and memory bandwidth, allowing the transformation pipeline to choose precision, layout, and buffering modes that maximise arithmetic intensity while adhering to accuracy constraints. Further investigations may also target hardware-specific optimisations aimed at bridging the performance gap between the hardware vendors. \added[id=Ours]{This includes studies on the impact of shared memory which allows GPU compute kernels to work without explicit data mapping.} By addressing these challenges, it may be possible to achieve more portable performance benefits across heterogeneous accelerator ecosystems.

%% file: 99_acknowledgements.tex
\added[id=Ours]{
Pawel's PhD studentship is partially supported by Intel's Academic Centre of
Excellence at Durham University.
This project has received funding through the UKRI Digital Research Infrastructure Programme under grant UKRI1801 (SHAREing) and grant number UKRI/ST/B000293/1 (HAI-End).
}

\added[id=Ours]{
The work has made use of the DiRAC@Durham facility managed by the Institute for Computational Cosmology
on behalf of the STFC DiRAC HPC Facility
(\href{www.dirac.ac.uk}{www.dirac.ac.uk}). The equipment was funded by BEIS capital funding via STFC capital grants ST/K00042X/1, ST/P002293/1, ST/R002371/1 and ST/S002502/1, Durham University and STFC operations grant ST/R000832/1. DiRAC is part of the National e-Infrastructure.
}

%% file: appendix_clang.tex
The forked Clang/LLVM project is publicly available at
\linebreak 
\url{https://github.com/pradt2/llvm-project}, our extensions are available in the \texttt{hpc-ext} branch.

The \texttt{hpc-ext} branch is up to date with the upstream LLVM \texttt{main} branch as of 12 Feb 2025.

To avoid conflicts, it is strongly recommended to remove any existing Clang/LLVM installations before proceeding.

To build and install the compiler toolchain, clone or otherwise download and unpack the repository, create a build folder in the top-level repository directory, and execute all steps from Algorithm~\ref{algorithm:appendix:compiler-download:build-compiler}.

\begin{algorithm}[htb]
    \begin{algorithmic}[1]
      \State 
      \begin{verbatim}
cmake 
  -DCMAKE_BUILD_TYPE="RelWithDebInfo" 
  -DCMAKE_C_COMPILER="gcc" 
  -DCMAKE_CXX_COMPILER="g++" 
  -DLLVM_ENABLE_PROJECTS="clang;lld" 
  -DLLVM_ENABLE_RUNTIMES="openmp;offload" 
  -DLLVM_TARGETS_TO_BUILD="X86;
                           AArch64;
                           NVPTX;
                           AMDGPU"
  ../llvm
      \end{verbatim}
\State \verb|make && make install|
    \end{algorithmic}
  \caption{
    Building and installing the compiler
    \label{algorithm:appendix:compiler-download:build-compiler}
    }
\end{algorithm}

\noindent

From hereon, \texttt{clang++} directs to our modified compiler version.


The command-line interface (CLI) is backwards compatible with the upstream Clang/LLVM version. The support for the new attributes discussed in this paper is enabled by default, no additional compiler flags are needed.

If the use of any of the new attributes leads to a compilation error, a common troubleshooting starting point is to inspect the rewritten source code. To see the rewritten code, add \texttt{-fpostprocessing-output-dump} to the compilation flags. This flag causes the post-processed source code be written to the standard output.